# STATE OF THE ART PARALLEL APPROACHES FOR RSA PUBLIC KEY BASED CRYPTOSYSTEM

Sapna Saxena and Bhanu Kapoor

Chitkara University, Himachal Pradesh, India

## ABSTRACT

*RSA is one of the most popular Public Key Cryptography based algorithm mainly used for digital signatures, encryption/decryption etc. It is based on the mathematical scheme of factorization of very large integers which is a compute-intensive process and takes very long time as well as power to perform. Several scientists are working throughout the world to increase the speedup and to decrease the power consumption of RSA algorithm while keeping the security of the algorithm intact. One popular technique which can be used to enhance the performance of RSA is parallel programming. In this paper we are presenting the survey of various parallel implementations of RSA algorithm involving variety of hardware and software implementations.*

## KEYWORDS

*Public Key Cryptography, RSA Algorithm, Security, Parallel Implementation.*

## 1.INTRODUCTION

Security has always been a biggest concern for the computing world in terms of transmitting information and data across the networks. Plenty of sensitive information that includes the data involved in electronic business, money transactions, etc. need to be transmitted securely over the Internet. Therefore, there has to be some technique that can ensure the security and integrity of the data to be transmitted on such un-trusted communication channels. Cryptography [1] is one of the most popular techniques that can be used to provide sufficient security to the sensitive data.

Cryptography is the art of disguising the data in such manner that only intended recipient can interpret the data. It is the combination of two different techniques – Encryption and Decryption. In encryption the message is converted to cipher text using encryption key and to decrypt the cipher text decryption key is used. For Encryption and Decryption various cryptographic algorithms are used that can be broadly classified into two distinct categories – Symmetric or Private Key Algorithms and Asymmetric or Public Key algorithms. Symmetric Key algorithms use a single key to encrypt as well as decrypt the data whereas Public key algorithms [2] use two distinct keys for encryption and decryption. The encryption key is called private key and the decryption key is called public key. RSA is the most popular public cryptographic algorithm.

Cryptographic algorithms are usually implemented serially that is each instruction of the program is executed one after other starting from the beginning towards end on a single processor. The public key based algorithms have their roots in modular arithmetic and they involve complex calculations involving very large integers. Due to the serial implementation of these compute-intensive algorithms slow down the process as it takes time to perform calculations for encryption

International Journal on Computational Sciences & Applications (IJCSA) Vol.5, No.1,February 2015

as well as decryption. It also requires large amount of memory which is sometimes not possible for a single processor. In addition to this due to the compute intensive behaviour they consume lot of power while executing themselves. By parallelizing these algorithms the power consumption can be reduced and high performance can be achieved in terms of time as well.

Parallel programming is an emerging area developed as a means of improving performance and efficiency which uses multi core processor for the faster and efficient execution of the instructions. In a parallel program, the processing is broken up into parts, each of which can be executed concurrently on different CPUs. Therefore to achieve higher performance in the area of security, the security algorithms can be implemented parallel in such manner that after dividing them in specific parts they can be executed on multi core processor to increase the speed and efficiency of it.

In the next sections basics of public cryptography and RSA algorithm is presented. Thereafter state of the art parallel approaches for RSA implementation are reviewed followed by the conclusion of the paper.

## 2.PUBLIC KEY BASED CRYPTOGRAPHIC ALGORITHMS

In public key based cryptographic algorithms two different keys are used to perform encryption and decryption. Due to this reason these are also called Asymmetric Cryptographic Algorithms. To carry out secure communication between two parties the receiver generates the public key and the corresponding private key using any public key based algorithm such as RSA. Then receiver shares the public key with the sender using any secured medium and requests him to initiate the communication. Now the sender encrypts the data using the public key and sends the cipher text to the receiver. On the other hand the receiver decrypts the cipher text using corresponding private key and gets the message.

There are many public key based algorithms are available. Some of the popular ones are – RSA Algorithm, Digital Signature Algorithm, Diffie-Hellman Key Exchange Algorithm, etc.

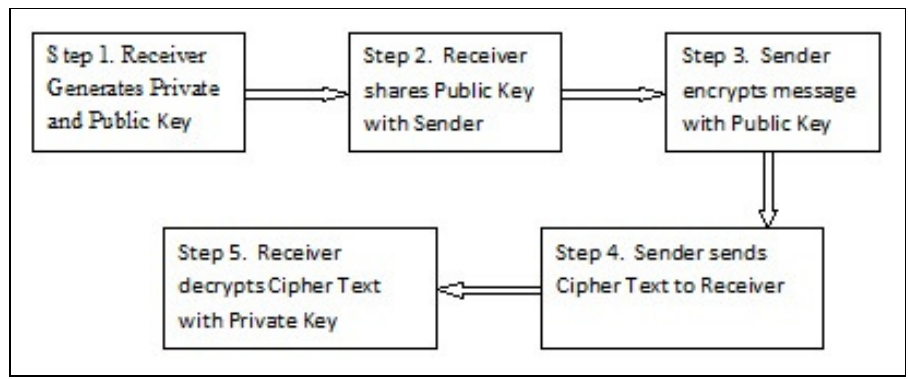

Figure 1: Mechanism of Public Key Cryptography

## 3.RSA ALGORITHM

The RSA algorithm [3] was invented by three MIT mathematicians Ronald Rivest, Adi Shamir and Leonard Adleman in 1978 and it is named after their names. RSA is used in many important areas such as for key exchange, to generate digital signatures and for the encryption and decryption, etc. RSA uses a variable key size which is usually taken as very large to ensure the security of the algorithm. RSA can be applied on variable size of encryption and decryption block.

RSA is based on the factorization of very large number, i.e. to find those two large prime numbers whose product is that number. Practically it is very difficult to find such numbers because usually the keys used in RSA are taken very large. Therefore the factorization takes very long time even using the best known algorithms This fact provides the tough security to the data subject to if provides with sufficiently long keys. For example, if the key length of 1024 bits or more is used then it is nearly impractical to break up the security of RSA encryption even when working with high performance computers.

The RSA can be divided into three algorithms - Key Generation Algorithm, Encryption Algorithm, and Decryption Algorithm.

### 3.1 Key Generation Algorithm

The key generation algorithm of RSA is a step-wise process which is as follows –

1. Choose p and q, two very large random prime integers with bit size at least 512 or more
2. Calculate modulus m as,
3. m = p*q
4. Calculate φ(n) as,
5. φ(n)= (p-1) (q-1)
6. Choose an integer e, 1 < e < φ(n) such that:
   GCD(e, φ(n)) = 1 (where GCD is greatest common denominator)
7. Calculate d, 1 < d < φ(n) such that:
   ed ≡ 1 (mod φ(n))

Here e is used as Encryption exponent and d is used as Decryption exponent therefore e and n are published as public key and d and n are secured as the private key.

### 3.2.RSA Encryption Algorithm

In RSA algorithm encryption can be applied on variable size of message block. Therefore data can be divided into the blocks of data using any padding scheme such as PCKS#1 and following procedure is applied to it –

$$C = M^e \% m$$

Where M is the message block and C is sent as the cipher text to the other party.

### 3.3.RSA Decryption

In order to decrypt the cipher text following procedure is applied to it –

$$M = C^e \% m$$

Where M is the original plain text and C is the Cipher text

# 4. STATE OF THE ART PARALLEL IMPLEMENTATIONS OF RSA ALGORITHM

The table 1 describes the various parallel approaches to implement RSA proposed by different researchers worldwide in order to achieve high performance and throughput in the area of public key cryptography. After that the brief description of each approach is explained.

Table 1: State of the art Parallel RSA approaches

| Author | Technique | Result |
|---|---|---|
| Laurent Imbert and Jean-Claude Bajard in 2002 | RNS implementation of RSA | High Throughput |
| Anshuman Rawat and Shabsi Walfish in 2003 | Parallel Signcryption Standard using RSA with PSEP | signcrypt small as well as long data |
| Mathieu Ciet et al in 2003 | FPGA Implementation of RSA with Residue Number Systems | Execution of 1024-bit RSA in less than 150 ms |
| S.H. Tang et al in 2003 | modular exponentiation on RSA on a Xilinx XC2V3000-6 device | Faster Decryption |
| Chia-Long Wu et al in 2004 | New algorithm faster than Savas-Tenca-Koc algorithm | Speedup of 1.06 to 2.75 |
| Qiang Liu et al in 2004 | VLSI implementation of the RSA | 7% increase in the throughput |
| Weng-Long Chang et al in 2005 | three new DNA-based algorithms | factoring two large prime numbers on DNA-based computer |
| Yu-Shiang Lin et al in 2006 | Algorithm based on the factoring of RSA-64 integer on GPUs using CUDA | speedup of 1197.5x as compared to any other CPU based algorithm |
| Hong Zhang et al in 2012 | Parallel implementation of RSA on GPGPU | Speedup of 45x as compared to multi-core CPU |
| Masumeh Damrudi and Norafida Ithnin in 2012 | Parallel RSA using Tree Structure | Fast Encryption |
| Sonam Mahajan and Maninder Singh in 2014 | Parallel RSA on GPU using CUDA Framework | High speed up even when using large prime numbers |
| Sapna Saxena and Bhanu Kapoor in 2014 | Parallel RSA based on Repeated square-and-multiply method | High speed up, approximately 5X as compared to the sequential RSA |

In 2002, Laurent Imbert and Jean-Claude Bajard presented "A full RNS implementation of RSA" [4]. This technique is serial in nature but mentioned here with the various parallel implementations of the algorithm because it gives high throughput. In this paper the authors proposed an efficient hardware implementation of RSA cryptosystem which is sequential in nature but gives high througput. It is based on Residue Number System to perform faster arithmetic. They presented new Montgomery Multiplication Algorithm in RNS which shows the efficiency with two implementations of RSA. The authors described two different techniques to implement RSA. First solution converts all the numbers into complete RNS and after computation it is converted back to binary numbers. Their second solution is completely based on RNS system and it never used numbers in the binary form.

Next Year in 2003, Anshuman Rawat and Shabsi Walfish presented "A Parallel Signcryption Standard using RSA with PSEP" [5]. They proposed standard for signcryption which is a combination of signature and encryption using the Probabilistic Signature and Encryption Padding (PSEP) scheme. They proposed the method to parallelize RSA that – first the short message is divided into two parts M1 and M2. Then Hash Code is generated for the message then it is concatenated with blocks. Thereafter the concatenated string is encoded and Label block is constructed. Then each block is processed parallel to obtain the speedup.This method can be used to signcrypt small as well as long data. Moreover they proposed that their technique can be used to improve the performance and security of RSA based key exchange protocols. However they described one drawback also that their scheme is improper for those applications where extremely low bandwidth is available and short messages must be exchanged.

In the same year 2003, Mathieu Ciet et al presented "Parallel FPGA Implementation of RSA with Residue Number Systems - Can side-channel threats be avoided?" [6]. In this paper they proposed the parallel algorithm to perform faster modular exponentiation. They used Montgomery multiplication algorithm based on Residue Number System to develop their framework which is finally synthesized on FPGA. They were able to execute 1024-bit RSA in less than 150 ms. For the parallel implementation the authors used a set of similar kind of parallel coprocessors which are independent to each other because they can perform intensive modular computations on pretty large numbers. Each cell is connected to 16-bit wide communication bus and they are managed by a multiplier controller containing the sequence of operations to perform one 512-bit multiplication of the 'square-and-multiply' algorithm. The combination of the control unit and the set of cells form a large multiplication processor. Which is further used to fed data for modular exponentiation.

S.H. Tang et al presented "Modular Exponentiation using Parallel Multipliers" in 2003 [7]. They proposed an FPGA semi-systolic implementation for the modular exponentiation algorithm which can be applied to RSA algorithm. In their architecture they utilized 18X18 multipliers on FPGA and employed carry save addition scheme. Further they described in their paper that this architecture can be used to perform 1024 bit modular exponentiation on RSA which operate at 90 MHz on a Xilinx XC2V3000-6 device. They performed 1024-bit RSA decryption in 0.66 ms with the Chinese Remainder Theorem using this architecture. In this paper the authors mentioned that to parallelize the algorithm a serial-parallel multiplier is used. They mentioned that for a standard serial n-digit multiplication requires $O(n^2)$ single digit multiplications. Therefore they designed their algorithm in serial-parallel hardware design where processing n digits in parallel increases the number of cycles linearly with n.

Thereafter in 2004, Chia-Long Wu et al presented "Fast Parallel Exponentiation Algorithm for RSA Public-Key Cryptosystem" [8]. In this paper they proposed an efficient technique for parallel computation of modular exponentiation where they get the speedup of 1.06 to 2.75. Basically they stressed upon the parallel computations of M E mod N where the bit-length of the numbers M, E, and N is taken as 512 to 1024 bits. They described in their paper that their algorithm is faster than Savas-Tenca-Koc algorithm in time complexity and it can improve the efficiency of RSA cryptosystem.

Qiang Liu et al presented "A regular RSA processor" in 2004 [9]. They proposed High performance VLSI implementation of the RSA algorithm based on the systolic array. To parallelize the RSA algorithm in addition to systolic array, they used a block-based scheme to eliminate global signals, with a pipelined bus to convey data globally. To calculate the iterative step of the algorithm they used carry-save-adder structure which improves the speedup of the algorithm. Further they proposed architectures to eliminate the fanout bottleneck. At the end they compared original modular multiplier architecture with fanout bottleneck and proved that their proposed architecture can provide 7% increase in the throughput without increase in area.

Next year in 2005, Weng-Long Chang et al presented "Fast Parallel Molecular Algorithms for DNA-Based Computation: Factoring Integers" [10]. They mentioned in their paper that the most complicated part of RSA algorithm is the factoring the product of two large prime numbers. To solve this problem they developed and proposed three new DNA-based algorithms which are parallel subtractor, parallel comparator, and parallel modular arithmetic. They further described that they are the first to demonstrate solution of solving the complex problem of factoring two large prime numbers on DNA-based computer.

Yu-Shiang Lin et al presented "Efficient Parallel RSA Decryption Algorithm for Many-core GPUs with CUDA" in 2006 [11]. In this paper, they proposed an efficient parallel RSA decryption algorithm GPFA, for GPUs using CUDA. In this paper authors suggested GPU-based Pollard's p-1 Factorization Algorithm, GPFA to accelerate the speed of factoring the large number N by using a GPU with CUDA. They subdivided the computations in the Pollard's p-1 Factorization into independent iterations which are further assigned to multiple threads in GPU to process. In the experiments GPFA used the inter-task parallelization technique to perform all computations. They tested their algorithm to find out the result of factoring a RSA-64 integer. And the experimental results of their algorithm show the speedup of 1197.5x as compared to any other CPU based algorithm. However they didn't test their algorithm for factoring RSA-128 or larger integers.

Six years later in 2012, Hong Zhang et al presented "Comparison and Analysis of GPGPU and Parallel Computing on Multi-Core CPU" [12]. They presented in this paper that RSA algorithm is very compute intensive and CPU is not suitable to perform the modular exponentiation part of it. However, GPU due to its high parallel processing power is more suitable to perform such operations. In this paper they implemented RSA algorithm on GPGPU and perform the comparative analysis between the results obtained from GPGPU and CPU. For the parallel implementation of RSA on GPU they used the method of threads and threads block. The computation part of the program is divided into threads that in turn composed the thread blocks. These threads are assigned with the same piece of work that work in the shared memory in a synchronized fashion. The threads are executed in the shared memory area to reduce the global memory access problem which considerably improves the performance of the algorithm. The results obtained by their experiments show that the GPGPU version of RSA algorithm gives 45x speedup as compared to its CPU counterpart.

In the same year 2012, Masumeh Damrudi and Norafida Ithnin presented "Parallel RSA encryption based on tree architecture" [13]. In this paper they applied parallel processing on RSA using tree structure. They proposed that by parallelizing RSA the speedup and the performance of RSA can be improved. They also present the state of the art methods that can be used to improve the speedup of the RSA while maintaining its security intact.

After two years in 2014, Sonam Mahajan and Maninder Singh presented "Analysis of RSA algorithm using GPU Programming" [14]. In their paper they described that the GPU as a co-processor of CPU can be used to implement massive parallelism. They designed parallel RSA algorithm for GPU using CUDA framework and tested for both small and large prime numbers. They described that their parallel RSA algorithm can be used to reduce the security threats due to the use of small prime numbers and to increase the speed of the algorithm. They used the stepwise process to parallelize the algorithm – first the key values are provide to the CPU, second CPU copies the values on CUDA enabled device by allocating the memory, third GPU kernel is invoked by CPU, fourth GPU processes the data using number of threads equal to the message length, fifth the data is transferred back to CPU from the GPU.

In the same year 2014, Sapna Saxena and Bhanu Kapoor presented "An efficient Parallel Algorithm for Secured Data Communications using RSA Public Key Cryptography Method". In

International Journal on Computational Sciences & Applications (IJCSA) Vol.5, No.1,February 2015

this paper authors proposed the new Parallel RSA algorithm based on repeated square-and-multiply method. To parallelize the proposed algorithm the authors mentioned that they decomposed the data using data decomposition technique and distributed the chunks of data among the available cores of the processor. They used static mapping technique to map the task operating upon the provided data to the cores of the processor. Then this data is computed in parallel using the cores of the processor and finally combined together to the get the final result. Moreover their algorithm is scalable in nature and thus adapts itself according to the number of cores available on the target machine. They used OpenMP API in the presence of GCC infrastructure on Linux platform to test their algorithm and got approximately 5X speedup as compare to its sequential counterpart.

## 6.CONCLUSION

Due to its roots in modular arithmetic based on very large numbers, RSA is considered to be slow algorithm. But it is still used in many important areas like SSL/TLS protocol, Digital signature, etc. therefore it is the point of attraction among the researchers working on public key cryptography. Throughout the world researchers are trying to increase the performance of RSA in terms of Time and Energy consumption. The table 1 includes some such important techniques which are followed by researchers. The aim of this paper is to bring into the notice of upcoming researchers regarding various parallel RSA implementation techniques which are already made.

## Authors

**Sapna Saxena**

Sapna Saxena, Research Scholar in the Department of CSE, Chitkara University, Himachal Pradesh, India. She has post graduated in Computer Applications (MCA) from AAIDU, Allahabad. She did her M.Tech. IT from KSOU. Presently she is pursuing Ph.D. from the Chitkara University, Himachal Pradesh. She is doing her research work in the area of Network security and Parallel Computing.

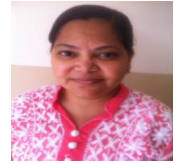

**Dr Bhanu Kapoor**

Dr. Bhanu Kapoor is the Professor in the Department of CSE, Chitkara University, Himachal Pradesh, India. He started his career with Texas Instruments where he played various technical roles (1987-99) at TI"s DSP R&D Center. He has played leading technology development roles in EDA startups ArchPro (now Synopsys), Atrenta, and Verisity (now Cadence). He is an expert in the area of low power design and verification. He is the lead inventor on 6 US patents in the area of low power design and verification and has over 40 publications in various IEEE/ACM conferences and journals.

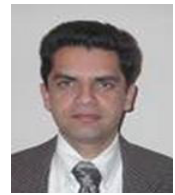